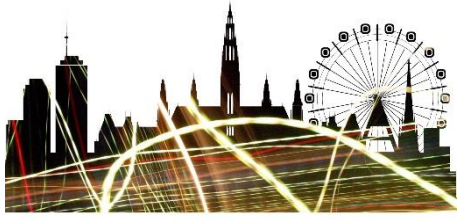

# *Towards an Interoperability Roadmap for the Energy Transition*

Valerie Reif, Florence School of Regulation, EUI, Italy, valerie.reif@eui.eu
Thomas I. Strasser, AIT Austrian Institute of Technology, Austria, thomas.strasser@ait.ac.at
Joseba Jimeno, Fundacion TECNALIA Research & Innovation, Spain, joseba.jimeno@tecnalia.com
Marjolaine Farre, TRIALOG SAS, France, marjolaine.farre@trialog.com
Olivier Genest, TRIALOG SAS, France, olivier.genest@trialog.com
Amélie Gyrard, TRIALOG SAS, France, amelie.gyrard@trialog.com
Mark McGranaghan, EPRI Europe, Ireland, mmcgranaghan@epri.com
Gianluca Lipari, EPRI Europe, Germany, glipari@epri.com
Johann Schütz, OFFIS e.V., Germany, johann.schuetz@offis.de
Mathias Uslar, OFFIS e.V., Germany, mathias.uslar@offis.de
Sebastian Vogel, E.DSO for Smart Grids, Belgium, sebastian.vogel@edsoforsmartgrids.eu
Arsim Bytyqi, ENTSO-E, Belgium, arsim.bytyqi@entsoe.eu
Rita Dornmair, B.A.U.M. Consult GmbH, Germany, r.dornmair@baumgroup.de
Andreas Corusa, B.A.U.M. Consult GmbH, Germany, a.corusa@baumgroup.de
Gaurav Roy, RWTH Aachen University, Germany, groy@eonerc.rwth-aachen.de
Ferdinanda Ponci, RWTH Aachen University, Germany, fponci@eonerc.rwth-aachen.de
Alberto Dognini, Fraunhofer FIT, Germany, alberto.dognini@fit.fraunhofer.de
Antonello Monti, Fraunhofer FIT, Germany, antonello.monti@fit.fraunhofer.de

**Abstract** – Smart grid interoperability is the means to achieve the twin green and digital transition but remains heterogeneous and fragmented to date. This work presents the first ideas and cornerstones of an *Interoperability Roadmap for the Energy Transition* that is being developed by the Horizon Europe int:net project. This roadmap builds on four cornerstones that address open interoperability issues. These are a knowledge base to address the lack of convergence among existing initiatives, a maturity model and a network of testing and certification facilities to address the lack of practical tools for the industry, and a governance process to address the gap between standards-related approaches of *Standards Development Organisations* and *Research and Innovation projects*. A community of practice will be set up to ensure the continuity of the ongoing activities related to smart grid interoperability. To outlive the duration of the int:net project, the aim is to formalise the community of practice as a legal entity.

## 1. Introduction

The European Union's (EU) 2030 and 2050 climate goals require the energy sector to undergo an unprecedented twin digital and green transition, which has already started. There is consensus that electricity generation from renewable resources, combined with a smart digital grid infrastructure, is the only way to decarbonise the energy sector. Final consumers and the overall demand side are playing an increasingly important role in this transition with central elements being easiness of access to data, transparency, and openness. The transition equally affects other sectors including mobility, buildings, and the Internet of Things (IoT) as outlined in the European Commission's (EC) *Energy System Integration Strategy*[1] and *Digitalisation of Energy Action Plan*[2].

Connectivity within and beyond the energy sector is needed to make the twin digital and green transition happen. Data exchange within the energy sector and with its interconnected domains, relying on Information and Communication Technology (ICT), is key to achieving a successful energy transition. However, there are sector-specific challenges due to certain energy system characteristics. Examples are the "systems-of-systems" nature of the energy system with legacy structures, traditional silo-thinking, and the use of proprietary instead of open standards. The increasing cross-sectoral focus of policymakers adds to the challenge to coordinate not one but different sectors at the same time. Complexity is added by the existence of diverse regulatory frameworks across the EU member states and the consensus that there is not one solution that fits all.

Interoperability is the means to overcome these challenges to achieve the transition. The future integrated energy system depends on the ability to exchange information across domains, organisations, actors and systems, and effectively use it to conduct mutual operations based on commonly agreed methods and processes in pursuit of joint objectives. While technical interoperability is quite well established in the energy sector, informational, functional, and business interoperability need more attention. [1], [2]

---

[1] https://eur-lex.europa.eu/legal-content/EN/ALL/?uri=COM:2020:299:FIN
[2] https://eur-lex.europa.eu/legal-content/EN/TXT/?uri=CELEX%3A52022DC0552&qid=1666369684560

In recent years numerous activities and projects related to smart grid interoperability have been carried out, producing valuable results, and defining important good practices and standards[3]. However, smart grid interoperability is still heterogeneous and fragmented.

This work presents the first ideas and cornerstones of the *Interoperability Roadmap for the Energy Transition* developed by the Horizon Europe *Interoperability Network for the Energy Transition (int:net)* project[4]. The main contribution of int:net is to bring all relevant initiatives and related actors together to efficiently and effectively advance interoperability for the EU twin green and digital transition.

The paper is structured as follows: Section 2 provides an overview of existing interoperability policy frameworks and initiatives and highlights remaining gaps and open issues. Section 3 presents the int:net cornerstones for an interoperability roadmap for Europe. Section 4 introduces the Interoperability Network for the Energy Transition. Section 5 concludes the work and provides an outlook on future plans.

## 2. Overview of EU Policy Framework and Initiatives for Interoperability

In the following, the existing EU policy framework for interoperability (cf. Section 2.1) as well as existing interoperability initiatives at the national, European, and international levels (cf. Section 2.2) are outlined. Finally, Section 2.3 highlights the remaining gaps and open issues.

### 2.1 Existing EU Policy Framework

The European policy framework for interoperability in the energy sector is still under development. Directive 2009/72/EC[5] of the *Third Energy Package* considered interoperability relevant in the context of an efficient operation and coordinated development of the interconnected electricity (transmission) system as well as regarding smart metering systems to be implemented across the member states. At the transmission level, ENTSO-E has driven the process towards increasing interoperability for the electricity market and network data, including through the development of common tools and processes [3]. At the distribution level, the smart grid interoperability landscape is much more fragmented, especially when it comes to the management of consumer data [4].

In 2011, the EC acted to enhance standardization activities and issued the Smart Grid Mandate M/490[6] to the European Standards Organizations (ESO) CEN, CENELEC, and ETSI. The ESOs were asked to develop a framework to identify standardization gaps, required use cases and security requirements in the field of smart grids, which resulted in the creation of the Smart Grid Architecture Model (SGAM) framework (see Section 2.2). Since then, the SGAM has been widely applied in European Research, Development, and Innovation (RD&I) projects including int:net.

---

[3] https://cinea.ec.europa.eu/publications/supporting-innovative-solutions-smart-grids-and-storage-2021_en
[4] https://intnet-project.eu
[5] https://eur-lex.europa.eu/legal-content/EN/ALL/?uri=celex%3A32009L0072
[6] https://ec.europa.eu/growth/tools-databases/mandates/index.cfm?fuseaction=search.detail&id=475

Directive (EU) 2019/944[7] of the *Clean Energy Package* considered interoperability of energy services in the context of smart metering as a prerequisite for customer empowerment and to promote competition in national electricity retail markets. The directive requires the member states to facilitate the full interoperability of energy services within the EU and enables the EC to adopt implementing acts. The acts will lay down interoperability requirements and non-discriminatory and transparent procedures for access to metering and consumption data and data required for customer switching, demand response and other services. The first of a series of acts is currently under development[8].

The *European Green Deal* has strengthened the focus on the cross-sectoral aspects of interoperability: The *Energy System Integration Strategy* stresses the importance of interoperability for a future decarbonised energy system integrated with the mobility and building sectors. A proposal for a recast of the *Energy Performance of Buildings Directive*[9] foresees future implementing acts regarding interoperability and access to building systems data (i.e., all data related to the energy performance of building elements, the energy performance of building services, building automation and control systems, and meters and charging points for e-mobility). In addition, the *European Data Strategy*[10] highlights the importance of interoperability for fostering data-driven innovation and creating a European single market for data, including the future European data spaces for energy and other sectors.

### 2.2 Existing Initiatives

The interoperability landscape for energy services is heterogeneous and wide; many initiatives exist at international, European, and national levels, covering various domains of the energy sector. This includes non-exhaustively research projects, best practices, working groups, ongoing standardisation activities, common frameworks, demonstrations, and policy initiatives.

*Research projects*, such as the ones funded by the EC under the Horizon 2020 or Horizon Europe programmes, gather experts from different sectors (including research institutes, small and medium-sized enterprises, universities, associations, and industry) to stimulate the development of knowledge and technologies deemed crucial for tackling challenges such as the digitalisation of the energy sector. On the topic of interoperability, several projects were funded. This includes the InterConnect project[11], focusing on developing and demonstrating advanced solutions for connecting and converging digital homes and buildings with the electricity sector, and the DRIMPAC project[12], focusing on the development of a unified Demand Response (DR) interoperability framework enabling market participation of active energy consumers. Other projects such as INTERRFACE[13] or OneNet[14] have made important contributions to the building of a replicable and scalable interoperable architecture to provide grid services for the efficient functioning of the power system. To capitalize on the lessons learned in each of these projects, the EC's BRIDGE[15] initiative groups projects and coordinates their work, including on interoperability issues.

---

[7] https://eur-lex.europa.eu/legal-content/EN/TXT/?uri=celex%3A32019L0944
[8] https://ec.europa.eu/info/law/better-regulation/have-your-say/initiatives/13200-Access-to-electricity-metering-and-consumption-data-requirements_en
[9] https://eur-lex.europa.eu/legal-content/EN/TXT/?uri=celex%3A52021PC0802
[10] https://eur-lex.europa.eu/legal-content/EN/TXT/?uri=celex%3A52020DC0066

Considering a top-down approach, Standards Development Organizations (SDO) such as IEC, CENELEC, and ISO and industrial alliances such as the Connectivity Standards Alliance develop, through dedicated working groups, *international standards* that represent a global consensus of state-of-the-art know-how. Such organizations can also propose *testing procedures and/or certification* to ensure the correct use of the standard by external actors (e.g., ENTSO-E's CGMES conformity assessment scheme or the OCPP 1.6/2.0 certification programmes proposed by the Open Charge Alliance).

*Common frameworks* are also developed for harmonization purposes between various stakeholders and countries. One example is the SGAM framework which provides a three-dimensional architectural framework that can be used to model interactions between different entities located within the smart energy area and which is widely used in the energy sector [5]. Another example is the *Harmonized Electricity Market Role Model*, a framework developed by ENTSO-E, EFET and ebIX to facilitate the dialogue between the market participants from different countries through the designation of a single name for each role and domain that are prevalent within the electricity market [6]. There are also common frameworks developed in other sectors that are relevant to the energy sector, for instance, the European Interoperability Framework (EIF)[16] in the public administration sector. The EIF considers four levels of interoperability: legal, organisational, semantic, and technical. The EIF helps to raise awareness and organise concepts and relevant terminology to facilitate the identification and prioritisation of interoperability issues. Based on the EIF, member states are required to develop national interoperability frameworks.

---

[11] https://interconnectproject.eu
[12] https://www.drimpac-h2020.eu
[13] http://www.interrface.eu
[14] https://onenet-project.eu
[15] https://bridge-smart-grid-storage-systems-digital-projects.ec.europa.eu
[16] https://joinup.ec.europa.eu/collection/nifo-national-interoperability-framework-observatory/european-interoperability-framework-detail

**2.3 Open Issues**

Despite the numerous existing interoperability initiatives, involving various actors from both the academic and industry ecosystem, several interoperability gaps and open issues exist. The following three have already been identified in the course of int:net.

The first open issue is a lack of coordination and convergence among the numerous ongoing interoperability initiatives and related actors. Where coordination exists, it is voluntary and can always only cover a subset of the relevant actors. The two following examples show the impact of uncoordinated action:

- Two or more initiatives may develop different protocols, models, or solutions for the same need, thereby defining incompatible processes for the same requirements and causing scattering and weakening of interoperability. A related example is the existence of seven competitive protocols for the same interface in the domain of flexibility trading [7].
- Two or more initiatives may also, despite their willingness to support and advance the ecosystem, end up developing competitive tools or frameworks, which further exacerbates the fragmentation of the solutions landscape. A related example is the use case repositories that have been developed by several actors including IEC, BRIDGE, EIRIE, and the JRC.

The second open issue is the lack of practical tools for industrial use. Indeed, while existing initiatives and frameworks excel in the provision of approaches and solutions to develop smart energy systems, they lack practical tools for the industry to

- assess the maturity of a system and its subsystems from an interoperability perspective;
- test and validate interoperability (incl. interoperability assurance).

Such tools and related practices would support system engineering to enable interoperability.

The third open issue is a gap between the approaches applied by RD&I activities, including EU-funded projects, and formal standardisation activities by SDOs. The former typically adopts a bottom-up approach, designing and developing solutions to demonstrate innovations, while the latter makes use of a top-down approach, willing to find a consensus among experts on how to proceed or solve a specific issue without experimental considerations. Therefore, a bidirectional collaboration between RD&I initiatives and formal standardisation activities is beneficial for all. RD&I projects should share novel requirements, test draft standards, and provide feedback to advance standards development. SDOs could benefit from the experience of RD&I projects and receive implementation feedback already during the standard development phase.

## 3. Main Cornerstones for an Interoperability Roadmap for Europe

In the following the four main cornerstones that form the int:net interoperability roadmap to address the open issues described in the previous section are addressed. They are a knowledge base to address the lack of convergence (cf. Section 3.1), a maturity model (cf. Section 3.2) and a testing and certification concept (cf. Section 3.3) to address the lack of tools for the industry as well as standards and governance guidance (cf. Section 3.4) to address the existing gap between RD&I and formal standardisation activities.

The cornerstones are embedded in the int:net approach as depicted in Figure 1. The int:net approach starts from the collection and assessment of existing frameworks and good examples (*Cornerstone 1: Knowledge Base*). These feed into the int:net maturity model (*Cornerstone 2: Maturity Model*). They provide means and measures for testing interoperability in a growing network of testbeds and testing laboratories (*Cornerstone 3: Testing and Certification*). Finally, they are promoted to framework setters to integrate them into their policy and support programmes (*Cornerstone 4: Standards and Governance*). Experiences from such tests and from setting up favourable frameworks will again feed into the knowledge base of good examples, which closes the virtuous interoperability circle.

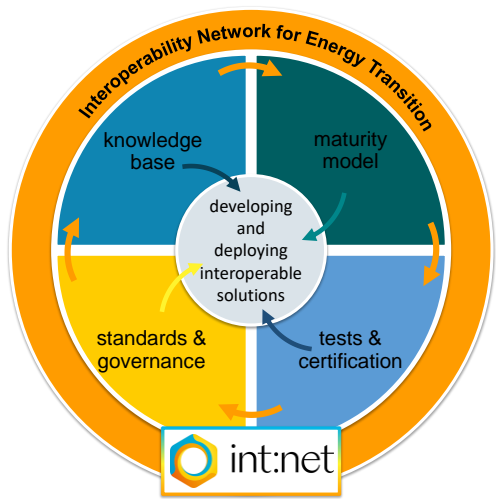

*Figure 1: Overview of the Interoperability Network for Energy Transition approach*

int:net will establish a community of practice to ensure the continuity of the ongoing interoperability activities and facilitate the coordination and alignment of relevant projects and initiatives happening at the different geographical levels (see Section 4). Without such a community there is a risk that interoperability expertise remains scattered among a few technical experts, redundant (and siloed) initiatives continue to exist, and changes in requirements, emerging use cases, test cases, and regulatory conditions are not adequately considered, hindering the full interoperability of energy services in a future integrated energy system. The establishment of the *Interoperability Network for Energy Transition* will ensure the sustainability of the int:net approach beyond the project's lifetime.

### 3.1 Common Knowledge Base for European Energy Services Interoperability

Following the FAIR (Findability, Accessibility, Interoperability, and Reusability) principles, int:net will establish and maintain a knowledge base of interoperability actions and best practices to increase interoperability of energy services, data, and platforms, both at the function and business layers. Relevant use cases from the already available repositories (incl. BRIDGE, ETIP SNET, EPRI, and the LOV4IoT-Energy ontology catalogue) will be analysed from the point of view of interoperability. In addition to this, the value chain of interoperability will be analysed by looking at business requirements and the associated business models. This analysis will be done both from the functional point of view (who specifies the technical features, who develops them, who maintains them) as well as from the economic point of view (who pays for development, validation, deployment, and maintenance of interoperability-related functions). Finally, the knowledge base will also consider the definition of the interoperability lifecycle process. This includes not only the consideration of interoperability during the system development phase but also during its entire lifecycle (operation, maintenance, upgrade, etc.).

Having a scope beyond a mere repository, the knowledge base will offer its users the option to collaborate on selected documents and existing institutions' initiatives. The knowledge base will be linked to existing platforms and include interoperability best practices, methodologies, and reference use cases coming from the frameworks of existing EU-wide and international initiatives. The platform will be supported through its growing community of registered users, which, together with the *Interoperability Network for Energy Transition*, will ensure that it is self-sustainable after the end of the project.

### 3.2 Comprehensive and Accepted Interoperability Maturity Model

Interoperability is a key requirement to take advantage of new technology innovations and integrate them as part of the overall energy transition. An interoperability maturity model and assessment procedure (reference implementation) will help organisations to understand their level of maturity in the integration of smart grid technologies, distributed generation, and customers with the planning and operation of grids and markets. The concept of maturity models has been used in this way previously for overall smart grid implementation [8], cyber security [9] and other cross-cutting technology areas. The int:net project builds on existing maturity model concepts to develop a model focused on the implementation of interoperability concepts and standards. This will highlight industry interoperability standards and architectures (mainly the IEC 63200 SGAM model, Levels of Conceptual Interoperability Model (LCIM), and SURF Green IT Maturity Model (SGIMM)) and will also highlight gaps where additional industry standards development and adoption are needed.

Maturity models involve the definition of domains and characteristics of maturity within each of the domains. Therefore, the initial effort in int:net is to define the domains to be considered for the Interoperability Maturity Model (IMM). Under each of these domains, specific characteristics that represent different levels of maturity need to be defined. These are typically divided into five levels of maturity, such as initiating, enabling, integrating, optimizing, and pioneering. Characterizing the maturity level in the different domains will help the industry identify priorities for ongoing improvements in interoperability standards and approaches.

Equally important, the project will develop a reference implementation of the model that will allow organizations to assess their progress in implementing important interoperability concepts. A workshop will help provide guidance and example applications of the maturity model framework. Documentation of the assessments over time assists the industry in tracking progress in supporting interoperability for accelerated adoption of new technologies and systems. Overall, feedback from the IMM application will accelerate the adoption and integration of new technology through continued improvement and understanding of the interoperability concepts and requirements.

### 3.3 Framework for Interoperability Testing in a Network of Testing Facilities

Interoperability is an essential aspect for the realisation of various applications and services in the domains of power and energy system where different approaches, solutions, components, and devices need to work together. There are several ways and principles to realise interoperability between two interconnected systems: by applying interoperability-by-design, by providing reference architectures as well as by following agreed rules, guidelines, and standards. To ensure interoperability in smart grids and energy systems it is not sufficient to follow these principles; instead, corresponding interoperability tests need to be realised. Certificates need to be issued, which provide evidence that the interoperability has been tested and can be ensured.

However, testing procedures, categorization, evaluation, and assessment criteria of the existing interoperability initiatives are very diverse and serve a wide range of interests. There is not yet a comprehensive overview of the existing interoperability testing approaches in the various smart grid and energy systems research infrastructures and laboratories across the EU member states. Also, despite the existence of many interoperability testing approaches, they are usually more advanced on the component/device level. On the system level, they are currently rather in the research and development stage. Moreover, existing approaches and related procedures and assessment criteria are diverse and serve a wide range of interests.

Therefore, int:net will establish a pan-European community of interoperability testing facilities. It also focuses on the harmonisation of existing testing procedures and concepts, the development, evolution, and integration of existing testing facilities as well as the formation of an integrated pan-European network of testing facilities and certification centres. A brandmark (work title "int:net approved") will be formally registered and an awarding procedure implemented.

Examples of relevant existing system-level based interoperability testing approaches are the Joint Research Center's (JRC) "Smart Grid Interoperability Testing Methodology" [10], the Austrian initiative "IES – Integrating the Energy System" [11], Interoperability Test "CIM for System Development and Operations" [12] and the "ERIGrid Holistic Test Description Approach" [13].

**3.4 Standards and Governance**

Standardisation usually drives the adaptation and utilisation of best practices for technical solutions and systems and fosters a shared understanding and (seamless) interoperability across independent vendors. Hence, standardisation promotes the global acceptance and competitiveness of solutions. Intending to prevent an arbitrary, unwanted growth of proprietary and not-interoperable solutions, adequate standardisation governance serves the general development, constant improvement and increase of acceptance of such standards by acting as a barrier-free, inclusive and trusted third party.

However, formal standardisation includes various bodies, domains, cross-cutting aspects and layers. Also, the number of standards that are relevant for critical infrastructures is very high. In addition, liaisons between groups exist, but standards evolve in parallel, thus, inconsistent development can sometimes not be avoided. Roadmaps issued by the IEC System Committees (SyC) prevent this but are typically limited to a particular domain, thus not considering cross-domain aspects, nor aligning standardisation efforts across domains.

In addition, regulation typically re-uses technical innovation and processes but alters them according to the respective national requirements. Therefore, standardisation per se does not lead to (technical) interoperability as has been shown with the example of smart metering in Europe. Over time, divide-and-conquer methods have been developed to distinguish and tackle various levels of interoperability. Typically, semantic, structural, and syntactic interoperability is focused on and solved based on the lowest level first, providing the seamless exchange of the payload to be interpreted later. Hence, also the standardisation process itself must adapt to those levels and structure them.

Therefore, int:net will develop a new governance process, test it, and promote it to governmental and regulatory institutions. Based on the IES experience [14], int:net will turn the SGAM and the existing maturity models into a practical process that brings together multiple stakeholders for co-creation sessions in so-called "Connectathons". Participants such as end-users, operators, product developers and vendors will cooperate to select and combine technical standards with market requirements in preparation for marketable, interoperable products. The results will be the basis for the formulation of policy recommendations and presented to programme managers and funding agencies.

## 4. Towards an Interoperability Network for the Energy Transition

To ensure the continuity of the ongoing activities related to the interoperability of energy services, data spaces and digital twins, the int:net solutions need to be publicly accessible, follow an open source and open science approach and stay up-to-date beyond the duration of the project itself. This will be ensured by establishing a community of practice, the *Interoperability Network for Energy Transition*.

The network will bring together all relevant actors, from the energy and related sectors, to work together on the development, testing, and deployment of interoperable energy services, laying the foundation for a future interoperable integrated energy system. A steady exchange between all stakeholder groups will ensure the implementation of best practices and common standards to address interoperability issues. The learnings will be leveraged by a diverse and open platform for interaction, collaboration, co-creation, and knowledge generation.

Starting from a well-balanced and connected consortium of researchers and framework setters including a national ministry and EU associations of Transmission System Operators (TSO) and Distribution System Operators (DSO), standardisation and communication experts, the int:net approach guarantees a wide outreach from the onset. To extend this core group of experts, int:net will engage with representatives of other initiatives such as working groups of the EC or SDOs, research projects, and the standardisation community that are working on common frameworks for testing interoperability at the European and national levels. int:net will leverage its access to existing communication channels through, among others, the ETIP SNET and BRIDGE platforms, the Joint Programming Platform Smart Energy Systems (JPP SES), IS-GAN, CENELEC, IEC, and Mission Innovation. The network will also reach out to industry alliances such as T&D Europe and AIOTI, and other initiatives such as BVDA, Gaia-X, OPEN DEI, IDSA, and the European Digital Innovation Centres.

Through the continuous exchange between interoperability initiatives and standardisation bodies, a broad consensus on the design of interoperable solutions will be achieved. The involvement of policymakers and regulatory authorities will enhance the understanding of how interoperability can be promoted and regulated at all levels. Eventually, the aim is to establish the network as a permanent legal entity, for example in the form of an association or non-profit organisation.

## 5. Conclusions

In recent years numerous activities and projects in the area of smart grid interoperability have been carried out. They have produced valuable results and defined best practices and standards. However, smart grid interoperability remains heterogeneous and fragmented. Therefore, this work presents the first ideas and cornerstones of an *Interoperability Roadmap for the Energy Transition* that is being developed by the Horizon Europe int:net project. The roadmap builds on four cornerstones that address open interoperability issues.

A knowledge base of interoperability best practices, methodologies and reference use cases will be created to address the lack of convergence among the numerous interoperability initiatives. The knowledge base will provide a comprehensive overview of relevant interoperability initiatives in the energy sector and beyond, at the national, European, and international levels.

An IMM and assessment procedure (reference implementation) as well as a testing and certification concept will be developed to address the lack of practical tools for the industry. The IMM will help to assess the level of interoperability maturity in organisations, promoting its adoption by other initiatives and uptake from the industry. Existing interoperability testing procedures and concepts will be harmonised and further developed, and a pan-European network of testing facilities and certification centres will be created.

A governance process for interoperability testing will be developed to address the existing gap between the approaches applied by RD&I activities, including EU-funded projects, and formal standardisation activities by SDOs. The process will also be promoted to governmental and regulatory institutions to ensure its uptake.

To ensure the continuity of the ongoing activities related to smart grid interoperability, a community of practice, the *Interoperability Network for Energy Transition,* will be established. This interdisciplinary network of stakeholders will collaborate to harmonise interoperability activities on energy services throughout Europe over the lifetime of the int:net project and beyond.

## Acknowledgements

This work has received funding from the European Union's Horizon Europe research and innovation programme under grant agreement N°101070086 (int:net).

## Authors

**Valerie Reif** is a research associate in electricity regulation at the Florence School of Regulation, European University Institute in Italy. She conducts research in European Horizon projects on interoperability and on TSO-DSO-consumer coordination. She teaches on electricity markets, the EU network codes, and the EU Green Deal. Valerie holds an MSc and a BSc degree in Renewable Energy Engineering and a BA in European Studies. She is a substitute member of the regulatory commission of the Austrian energy regulatory authority E-Control.

**Thomas I. Strasser** received a master's and a PhD degree from the Technische Universität Wien (TU Wien) and he was awarded the Venia Docendi (habilitation) in the field of automation from the same university. For several years, he has been a senior scientist in the Center for Energy of the AIT Austrian Institute of Technology. His main responsibilities involve the strategic development of smart grid automation and validation research projects as well as the mentoring/supervising of junior scientists.

**Joseba Jimeno** is B. Sc. in Industrial Engineering and M. Sc. in Electric Engineering (University of the Basque Country, 2001). He is working in Smart Grid related research projects as research engineer and project manager in TECNALIA. His main research activities are related to energy management in microgrids, demand response and the operation of distribution grids. He has been also involved in the development of communication architectures and information models for Smart Grid applications.

**Marjolaine Farre** holds a Diploma Engineer degree from Supélec and a Master's degree in Environmental Systems Engineering from University College London (UCL). In 2017, she worked as a consultant for Enedis, the main Distribution System Operator (DSO) in France, to improve their modelling tools for real-time operation of the MV and LV grids. In 2020, she joined Trialog to consolidate the Energy team and is now involved in several H2020 European projects related to smart grid, energy management, flexibility and interoperability.

**Oliver Genest** has 14-year experience in ICT for smart energy systems, focusing on system architecture and digital transformation of energy systems (interoperability, standards, IoT, data management, data spaces). At European level, he is the Chairperson of the BRIDGE working group on Data Management. Within Trialog, he leads the energy-related business and projects, and coordinates research and innovation activities. He holds an engineering degree from Ecole des Mines de Nancy, France, and a General Management certificate from ESCP Europe.

**Amélie Gyrard** is a R&I European project consultant at Trialog, Paris, France. She has 10-year experience in IoT Semantic Interoperability and experienced in working with H2020 European projects such as StandICT.eu 2023, AI4EU, Interconnect, and FIESTA-IoT etc. She co-authored white papers targeting developers and engineers where standardization activities are collaborating (W3C Web of Things, ISO/IEC JTC1, ETSI, ONEM2M, and AIOTI). She is involved in standardizations (e.g., IEC SyC Smart Energy, ISO/IEC SC41 IoT and Digital Twin, ISO/IEC SC42 AI). She co-authored more than 40 scientific articles (>1900 citations, h-index=29); and is a reviewer for communities such as IoT, WWW, and AI.

**Mark McGranaghan** is a Fellow at EPRI, located at the EPRI Europe office in Dublin, Ireland. He received his BS and MS degrees in Electrical Engineering from the University of Toledo (Ohio) in 1977 and 1978, respectively. He has authored more than 70 technical papers and articles on topics ranging from power quality to insulation coordination. He has been a leader in the development of smart grids for the last 20 years. He is an IEEE Fellow and in 2014 received the Charles Proteus Steinmetz Award for his expertise and dedication to power engineering standards development.

**Gianluca Lipari** received his Ph.D. degree in electronic engineering from the University of Reggio Calabria, Italy, in 2016. In 2015, he joined the Institute for Automation of Complex Power Systems of RWTH Aachen University in Germany and, in November 2020, the Fraunhofer FIT Center for Digital Energy in Aachen. Since October 2022 he is Technical Leader and European Project Manager at EPRI Europe. His research activities focus on digitalization of the energy system, including cloud applications for cyber-physical systems monitoring and automation.

**Johann Schütz** is a researcher in the field of Energy Informatics at OFFIS. He received his Bachelor and Masters Degree in Business Informatics at the University of Applied Sciences of Osnabrück in 2014 and the University of Oldenburg in 2017, respectively. After his graduation he worked at the OFFIS Institute for Information Technology in the group "Standardized Systems Engineering and Assessment" and is currently doing his Ph.D. with a focus on the digital sustainability of the energy system as a (net-centric) system-of-systems.

**Mathias Uslar** is member for the German NC in the IEC SyC Smart Energy WG 5 as well as in the various national German mirrors. He is senior principal scientist as well as Group Manager at the OFFIS – Institute for Information Technology in Oldenburg, Germany. His work focuses on the topic of Systems Engineering and Assessment, mainly focusing on the aspects of System-of-systems interoperability as well as IT security.

**Sebastian Vogel** represents E.DSO for Smart Grids in several EU-funded initiatives revolving around the digitalisation of distribution grids and cross-sector integration. He graduated from Fudan University, China, with an M.Sc in International Public Policy and from Gothenburg University, Sweden, with an M.Sc in International Administration and Global Governance. Sebastian connects technological innovation and European policymaking. He advocates for empowering stakeholders and collaborative outcomes to support the green transition and the digitalisation of power systems.

**Arsim Bytyqi** received his B.E. and M.Sc. degrees from University of Pristina, Pristina, Kosovo, in 2005 and 2009, respectively. He received his Ph.D. degree in Nanotechnology from Jozef Stefan International Postgraduate School, Ljubljana, Slovenia, in 2013. He is currently working as Advisor in ENTSO-E. He is a member of IEC TC57 WG13. His research interests include modelling of power grid in CIM/CGMES format, advancing standardization and interoperability activities in TSOs, and contributing on development and implementation of EU H2020 projects.

**Rita Dornmair** is researcher and consultant at B.A.U.M. Consult. In international research projects, she works on flexibility in distribution grids, smart energy, energy communities and interoperability in energy systems. In accompanying research on research programs, she develops methods for evaluating project performance as well as impact on entire systems. Rita Dornmair holds a Dr.-Ing. in the field of energy systems/energy economics and a Dipl.-Ing. in Electrical Engineering and Information Technology from the Technical University of Munich.

**Andreas Corusa** is consultant and researcher at B.A.U.M. Consult. In research projects and programs, he works on management and further development of digital collaboration and community platforms. He is involved in customer and citizen engagement processes and analyses in international research projects with focus on energy transition and digitization of energy systems. Andreas Corusa holds an MSc and BEng degree in Energy and Building Services Engineering and has several years of experience as Co-Founder, Entrepreneur and Consultant in Asia.

**Gaurav Roy** received the B.Tech. Degree from SRM University, Chennai, India, and the M.Sc. Degree in power engineering from RWTH Aachen University, Aachen, Germany, where he is pursuing a PhD degree with the Institute for Automation of Complex Power Systems (ACS).
He joined the Institute for Automation of Complex Power Systems (ACS), RWTH Aachen University, as a Research Assistant. His research topic is the automation of multi-terminal dc grids and the interoperability of devices in automation systems.

**Ferdinanda Ponci** received her Ph.D. degree in electrical engineering from the Politecnico di Milano, Italy, in 2002. She was with the Department of Electrical Engineering, University of South Carolina, Columbia, SC, USA, until 2008 and from 2009 with the Institute for Automation of Complex Power Systems, E.ON Research Center, RWTH Aachen University, Aachen, Germany, where she is currently a Professor for Monitoring and Distributed Control for Power Systems. Her research interests include advanced measurement, monitoring and automation of active distribution systems.

**Alberto Dognini** received his M.Sc. degree in electrical engineering from the Politecnico di Milano, Italy, in 2014. In 2015 he joined ABB - Medium Voltage Products Division as Engineering Project Manager. Since 2017 he has been a Research Associate with the Institute for Automation of Complex Power Systems, E.ON Energy Research Center at RWTH Aachen University, Germany and, since 2022, also with the Center for Digital Energy, Fraunhofer Institute for Applied Information Technology (FIT), Germany.

**Antonello Monti** received his Ph.D. in electrical engineering from the Politecnico di Milano, Italy, in 1994. He started his career in Ansaldo Industria and then moved to the Politecnico di Milano, as an Assistant Professor. In 2000, he joined the Department of Electrical Engineering of the University of South Carolina (USA), as Associate and then Full Professor. Since 2008, he is the Director of the Institute for Automation of Complex Power System at the E.ON Energy Research Center, RWTH Aachen University. Since 2019, he holds a double appointment with Fraunhofer FIT, developing the new Center for Digital Energy, Aachen.